# A NEW MIDDLE PATH APPROACH FOR ALIGNMENTS IN BLAST

DEEPAK GARG
*Thapar Institute of Engineering & Technology, Patiala, India*
dgarg@tiet.ac.in, deep108@yahoo.com

SURESH C SAXENA
Thapar Institute of Engineering & Technology, Patiala, India

LALIT M BHARDWAJ
Central Scientific Instruments Organization, Chandigarh, India

**Abstract**

This paper deals with a new middle path approach developed for reducing alignment calculations in BLAST algorithm. This is a new step which is introduced in BLAST algorithm in between the ungapped and gapped alignments. This step of middle path approach between the ungapped and gapped alignments reduces the number of sequences going for gapped alignment. This results in the improvement in speed for alignment up to 30 percent.

## 1. Introduction

In the plethora of tools available for data mining in bioinformatics, Basic Local Alignment Search Tool (BLAST) is being extensively used due to its unmatched speed and sensitivity. Though the performance of BLAST is the best in its class of tools but still there is a lot of scope of improvement in it. In order to work upon BLAST, its variants are understood and a lot of parameters, on which the speed and sensitivity of BLAST depends, are analyzed. The amount of research, which has gone so far into the BLAST, is tremendous. Many people have put in years of efforts to formulate the core of BLAST. In this work, an attempt has been made to improve its performance by taking into consideration its parameters and working of BLAST. There are various parameters that have contextual relations with areas other then the algorithm design and theory of computer science. However, in the present work, the analysis of these parameters has been limited from the viewpoint of a computer engineer. Due to increased traffic, BLAST is becoming slower and slower day by day. Also the number and size of sequences are increasing. Therefore, there is a need to continuously improve BLAST algorithm to keep its speed with the requirements of biologists. Even after a number of improvements in hardware and in parallel and distributed algorithms, BLAST is predicted to run to half of its speed every year. If this trend continue, then after 3-4 years it may not be possible to work with BLAST.

BLAST is a set of similarity search programs designed to explore all available DNA and protein sequence databases. BLAST programs have been designed for speed, with a minimal sacrifice of sensitivity to distant sequence relationships. BLAST uses the concept of a "segment pair" which is a pair of sub-sequences of the same length that form an ungapped alignment. The algorithm first looks for short words that are present in both the sequences and then extend these at either ends to find the longest segments present in both. The statistical significance of these High-scoring Segment Pairs is evaluated to determine whether the matches are random or not. Thus, the scores assigned in a BLAST search have a well-defined statistical interpretation, making real matches easier to distinguish from random background.

However, as how the sequences are classified and their functionality is found have changed with time. The molecular biologists are changing their approach with the advent of new techniques and technologies. When a threshold is crossed while finding similarity between two sequences, then sequences are said to be homologous. The Homology values have utility of finding biological properties, chemical properties and other characteristics of the sequence for which these were unknown previously. Normally everyone is using Smith-Waterman algorithm to find out local alignments. Till now it is considered as a good model to show similarities between two regions of two sequences with allowed number of mutations or differences or mismatches. [1-3]

The algorithms that were used to find out similarities have improved on time. First algorithm that became popular was FASTA [4]. Then BLAST [5] was the major entry in this area and till now it is very popular amongst scientists. There are many improvements that appeared in BLAST from time to time. The improvement can be in the number of hits and can be implemented in a multi hit algorithm that can actually take the value of N that should be used for the N-hit algorithm [6]. Also there can be a drop off percentage score instead of drop off score so that for calculating the drop off there is no need to go into the scoring matrix [7].

## 2. BLAST

In Lipman[8] algorithm, two sequences called as target and query sequence are compared. For this, it requires a matrix of size m X n; if the size of the sequences is m and n respectively. So as per algorithmic techniques, it will have $n^2$ complexity. This algorithm looks for overlapping regions of similarity of length W that is known as high scoring regions. The choice of value for W has a direct affect on the number of hits being produced. The value of W is a tradeoff in the speed and sensitivity. Small W gives more results. Every hit has its starting and finishing index in sequences being compared. Then the algorithm performs an ungapped extension if two hits are on same diagonal and the difference between starting indices is less then a constant A. For doing ungapped extension and to know that whether it results in a high Scoring pair alignment dynamic programming technique is implemented using a two dimensional matrix in which query sequence is on one side of the matrix and the target sequence is on the other side of the matrix [9].

Once the high scoring area is found out for ungapped region, then there is a need to go for gapped alignment to know whether better results can be found in the form of bigger area alignments. Here, a seed value is taken that has to be from the ungapped region. Then it tries to extend the matches towards both regions of the seed value. Here the drop-off parameter is used. The extension continues until the score does not falls below the drop off parameter and satisfies the eligibility criterion for the sequences to be displayed in result. The selection of the value of drop-off parameter is again the tradeoff between speed and sensitivity.

The highest score for any alignment is calculated. The score [a+1,b+1] is dependent upon the three cells [a,b], [a+1,b] and [a,b+1]. Similarly score [a+2,b+2] is dependent on [a+1,b+1], [a+2, b+1] and [a+1, b+2] . This can be extended similarly for an m X n matrix depending on the size of the target and query sequence. So 'a' lies from 1 to m and 'b' from 1 to n.

After this, depending upon the E-Value, nominal Score S2 and maximum number of sequences to be displayed by the user, the results are displayed. E- Value is a statistical parameter to find out the probability of finding the same score or higher score if the same query was searched against a random database [8-10].

$E = Q/2^{s'}$

Where $s' = (\lambda s - \ln K) / (\ln 2)$ and $Q = mXn$

m= total length of query sequence

n = total length of target sequence

$\lambda$ and K are constants and also depends on the scoring matrix e.g. BLOSUM or PAM

## 3. Middle Path Approach

The meaning of middle path approach is that instead of performing ungapped alignments as well as gapped alignments on all the sequences, only relevant sequences may pass through these two phases of algorithm. It is not necessary that all those sequences which are going for ungapped analysis may require gapped analysis. A middle path should be taken between ungapped and gapped alignments. The reason is that there is no need to perform gapped alignment on all the sequences on which ungapped analysis is performed. As per suggested approach, only some of the sequences should be sent for gapped alignment. If the algorithm of BLAST is studied in detail, it is found out that out of the total time one third is taken to find out word hits, one third is taken to find out ungapped alignment and one third is taken for gapped alignment. If ungapped analysis is performed for 100 alignments, then approximately one alignment crosses the eligibility mark to be displayed as a result of the sequence alignment.[13,14] So if a check can be made on a number of alignments that go from ungapped alignments to gapped alignments; and gapped alignment is performed only on these reduced number of sequences then time is reduced for gapped alignment and results in significant time saving. The time is reduced by that proportion by which there is reduction in ungapped alignments passing for gapped alignment. So a parameter is to be introduced as a check for making the sequences eligible for gapped alignment. For this, the understanding of how gap costs are calculated in BLAST is crucial.

### 3.1 Calculating affine gap costs

In this, the observation is that the insertion cost is high as compared to other costs like initiating a gap, extending a gap, because insertion cost is the sum of initiating a gap and insertion into that gap.

The recursive algorithm that uses dynamic programming and calculates the values for every cell is as follows

1. Best (a,b)=tempBest(a-1,b-1)+score[a,b]

Where score[a,b] is the original value in the 2-D matrix taken from dynamic programming matrix prepared using scoring matrices like PAM, BLOSUM or others and tempBest(a,b) is the best alignment score up to a certain point on the matrix diagonal.

Insertion$^q$ (a-1, b) means inserting at [a,b] with respect to target sequence and Insertion$^t$(a,b-1) means Inserting at [a,b] with respect to query sequence.

If (Insertion$^q$ (a-1, b) > Insertion$^t$(a,b-1))
{
if (Insertion$^q$ (a-1,b) > best(a,b)

```
            Insertion^q (a-1, b) = Insertion^q (a-1, b) – gap_extension;
    Else
            {
            Insertion^q (a-1, b) = Best (a,b) – Insertion_Cost;
            tempBest = Insertion^q (a-1,b) ;
            }
    Else If (Insertion^t(a,b-1) > Best(a,b)) then
            {
            Insertion^t(a,b-1) = Insertion^t(a,b-1) – gap_extension;
            tempBest = Insertion^t(a,b-1);
            }
    Else
            Insertion^t(a,b-1) = Best(a,b) – Insertion_Cost;
Else
    tempBest(a,b) = Best(a,b);
            }
```

The above algorithm makes only three comparisons as compared to five in the original BLAST algorithm and four in the algorithm suggested by Zhang, Pearson and Miller[16]. This helps in reduction in time. Array access can be done once for a particular value by assigning the value to a variable. That further helps in time reduction. The above algorithm makes only three arithmetic operations as compared to five in the original BLAST algorithm and four in the algorithm suggested by Zhang, Pearson and Miller.

Going further, there is a need to know when any insertion is to be made. For this, a strategy has to be got evolved as which one of the sequences has to go in for gapped alignment category. The parameter for deciding insertions and deletions is to be dependent on the ratio of the gapped and ungapped alignments. It has been observed that when the number of sequences travel from ungapped to gapped alignment, the resulting alignments are reduced to 1 to 2 %. So taking a clue from it, a lot of testing was done and thereafter it was observed that the insertions can be made at every N character where N can vary from 2 to 100. Here the value of N will make a balance between gapped and ungapped alignments. It was observed that higher the value of N, lower are the number of insertions and lesser are the calculations resulting in less speed. So the value of N will again make a tradeoff between speed and sensitivity. Unusual high value of N may result in loss of sensitivity.

Table 1. Indexes of the two dimensional matrix used to calculate the best score for gapped alignment

|          | Query sequence |           |           |     |
|----------|----------------|-----------|-----------|-----|
| Target   | [a,b]          | [a+1,b]   | [a+2,b]   | ... |
| sequence | [a,b+1)        | [a+1, b+1]| [a+2, b+1]| ... |
|          | [a, b+2]       | [a+1,b+2] | [a+2,b+2] | ... |
|          | ...            | ...       | ...       | ... |

The basis for above is that generally gaps are seen in the sequences in the regions that are less conserved [13]. It was also observed that it is not important whether the gap is at the start or at the end or somewhere else. Sometimes the gaps in the sequences are very longer and force the change of diagonal in the matrix. In some cases, the ungapped regions can be shifted if the gaps are unusually longer.

So depending upon whether both insertions Insertion$^q$ (a,b) and Insertion$^t$(a,b) are allowed or Insertion$^q$ (a,b) is allowed or only Insertion$^t$(a,b) is allowed or no insertion is allowed at that point, this will make an interesting change in proposed recursive algorithm as per the following

    Case 1:
    // Recursion step follows (No comparisons)
    Best (a,b)=tempBest(a-1,b-1)+score[a,b]
    Case 2:
    // Recursion step follows ( 1 comparison)
    Best (a,b)=tempBest(a-1,b-1)+score[a,b]
    If Insertion$^t$(a,b-1) > Best(a,b) then
        {
        Insertion$^t$(a,b-1) = Insertion$^t$(a,b-1) – gap_extension;
        tempBest = Insertion$^t$(a,b-1);
        }
    Else
        {
        Insertion$^t$(a,b-1) = Best(a,b) – Insertion_Cost;
        tempBest(a,b) = Best(a,b);
        }
Case 3:
    // Recursion step follows (1 comparison)
    Best (a,b)=tempBest(a-1,b-1)+score[a,b]
    if (Insertion$^q$ (a-1,b) > best(a,b)
        {
        Insertion$^q$ (a-1, b) = Insertion$^q$ (a-1, b) – gap_extension;
        tempBest = Insertion$^q$ (a-1,b) ;
        }
    Else
        {
        Insertion$^q$ (a-1, b) = Best (a,b) – Insertion_Cost;
        tempBest(a,b) = Best(a,b)
        }
Case 4:
    // recursion step follows (3 comparisons)
    Best (a,b)=tempBest(a-1,b-1)+score[a,b]
    If (Insertion$^q$ (a-1, b) > Insertion$^t$(a,b-1))
        {
    if (Insertion$^q$ (a-1,b) > best(a,b)
        Insertion$^q$ (a-1,b)= Insertion$^q$ (a-1,b) – gap_extension;
    Else
        {
        Insertion$^q$ (a-1,b) = Best(a,b) – Insertion_Cost;)
        tempBest = Insertion$^q$ (a-1,b) ;
        }

```
              Else If Insertion^t(a,b-1) > Best(a,b) then
                  {
                  Insertion^t(a,b-1) = Insertion^t(a,b-1) – gap_extension;
                  tempBest = Insertion^t(a,b-1);
                  }
              Else
                  Insertion^t(a,b-1) = Best(a,b) – Insertion_Cost;
                  }
          Else
              tempBest(a,b) = Best(a,b);
              }
```

| | | \multicolumn{19}{c|}{Target Sequence} |
| --- |---|---|---|---|---|---|---|---|---|---|---|---|---|---|---|---|---|---|---|---|
| | | 1 | 2 | 3 | 4 | 5 | 6 | 7 | 8 | 9 | 1 | 1 | 1 | 1 | 1 | 1 | 1 | 1 | 1 | 2 |
| Query Sequence | 1 | ↘ | ↘ | ↘ | ↘ | ↘ | ↘ | ↘ | ↘ | ↘ | ↓ | ↘ | ↘ | ↘ | ↘ | ↘ | ↘ | ↘ | ↘ | ↓ |
| | 2 | ↘ | ↘ | ↘ | ↘ | ↘ | ↘ | ↘ | ↘ | ↘ | ↓ | ↘ | ↘ | ↘ | ↘ | ↘ | ↘ | ↘ | ↘ | ↓ |
| | 3 | ↘ | ↘ | ↘ | ↘ | ↘ | ↘ | ↘ | ↘ | ↘ | ↓ | ↘ | ↘ | ↘ | ↘ | ↘ | ↘ | ↘ | ↘ | ↓ |
| | 4 | ↘ | ↘ | ↘ | ↘ | ↘ | ↘ | ↘ | ↘ | ↘ | ↓ | ↘ | ↘ | ↘ | ↘ | ↘ | ↘ | ↘ | ↘ | ↓ |
| | 5 | ↘ | ↘ | ↘ | ↘ | ↘ | ↘ | ↘ | ↘ | ↘ | ↓ | ↘ | ↘ | ↘ | ↘ | ↘ | ↘ | ↘ | ↘ | ↓ |
| | 6 | ↘ | ↘ | ↘ | ↘ | ↘ | ↘ | ↘ | ↘ | ↘ | ↓ | ↘ | ↘ | ↘ | ↘ | ↘ | ↘ | ↘ | ↘ | ↓ |
| | 7 | ↘ | ↘ | ↘ | ↘ | ↘ | ↘ | ↘ | ↘ | ↘ | ↓ | ↘ | ↘ | ↘ | ↘ | ↘ | ↘ | ↘ | ↘ | ↓ |
| | 8 | ↘ | ↘ | ↘ | ↘ | ↘ | ↘ | ↘ | ↘ | ↘ | ↓ | ↘ | ↘ | ↘ | ↘ | ↘ | ↘ | ↘ | ↘ | ↓ |
| | 9 | ↘ | ↘ | ↘ | ↘ | ↘ | ↘ | ↘ | ↘ | ↘ | ↓ | ↘ | ↘ | ↘ | ↘ | ↘ | ↘ | ↘ | ↘ | ↓ |
| | 1 | → | → | → | → | → | → | → | → | → | ↓ | → | → | → | → | → | → | → | → | ↓ |
| | 1 | ↘ | ↘ | ↘ | ↘ | ↘ | ↘ | ↘ | ↘ | ↘ | ↓ | ↘ | ↘ | ↘ | ↘ | ↘ | ↘ | ↘ | ↘ | ↓ |
| | 1 | ↘ | ↘ | ↘ | ↘ | ↘ | ↘ | ↘ | ↘ | ↘ | ↓ | ↘ | ↘ | ↘ | ↘ | ↘ | ↘ | ↘ | ↘ | ↓ |
| | 1 | ↘ | ↘ | ↘ | ↘ | ↘ | ↘ | ↘ | ↘ | ↘ | ↓ | ↘ | ↘ | ↘ | ↘ | ↘ | ↘ | ↘ | ↘ | ↓ |
| | 1 | ↘ | ↘ | ↘ | ↘ | ↘ | ↘ | ↘ | ↘ | ↘ | ↓ | ↘ | ↘ | ↘ | ↘ | ↘ | ↘ | ↘ | ↘ | ↓ |
| | 1 | ↘ | ↘ | ↘ | ↘ | ↘ | ↘ | ↘ | ↘ | ↘ | ↓ | ↘ | ↘ | ↘ | ↘ | ↘ | ↘ | ↘ | ↘ | ↓ |
| | 1 | ↘ | ↘ | ↘ | ↘ | ↘ | ↘ | ↘ | ↘ | ↘ | ↓ | ↘ | ↘ | ↘ | ↘ | ↘ | ↘ | ↘ | ↘ | ↓ |
| | 1 | ↘ | ↘ | ↘ | ↘ | ↘ | ↘ | ↘ | ↘ | ↘ | ↓ | ↘ | ↘ | ↘ | ↘ | ↘ | ↘ | ↘ | ↘ | ↓ |
| | 1 | ↘ | ↘ | ↘ | ↘ | ↘ | ↘ | ↘ | ↘ | ↘ | ↓ | ↘ | ↘ | ↘ | ↘ | ↘ | ↘ | ↘ | ↘ | ↓ |
| | 2 | → | → | → | → | → | → | → | → | → | ↓ | → | → | → | → | → | → | → | → | ↓ |

Fig.1. Dynamic programming matrix highlighting insertions in the target and query sequence that will be required for the middle path approach for 20X20 matrix

If the matrix size is aXb, then case 4 runs approximately $(aXb)/N^2$ times. For example, in the given matrix of 20X20, N =10, the case 1 runs $20X20/(10)^2$ i.e. 4 times. So the case that is most computing intensive out of the four runs least number of times. Case 1 runs approximately $(N-1)^2*((aXb)/N^2)$. For example, given in the matrix of 20X20, N=10, it runs for 81*4= 324 times. So the case with least amount of computation runs most of times. This effect is in same proportion for arithmetic instructions also. The combined effect makes significant improvement in the speed.

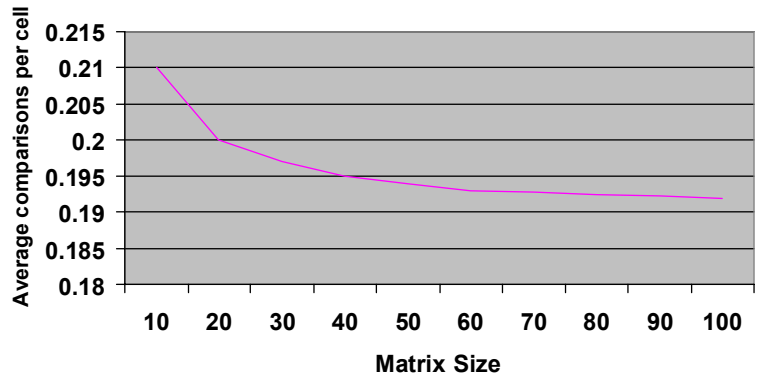

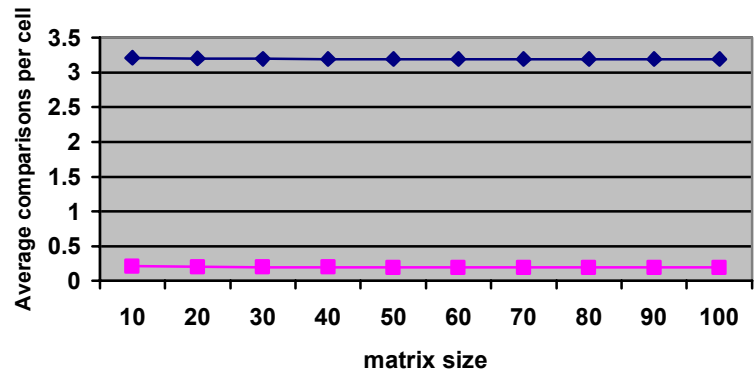

Fig.2a. Number of average comparisons required per cell if the matrix size is increased from 10 to 100 with same value of N, Data shows values for N=10.
Fig.2b. Comparison with average number of comparisons per cell in existing algorithm

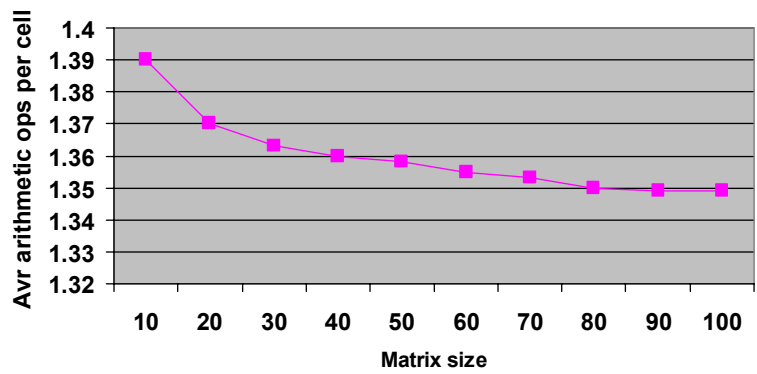

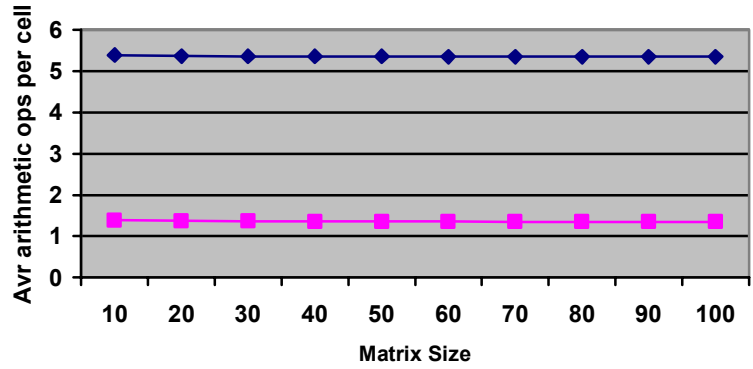

Fig.3a. Number of average arithmetic operations required per cell if the matrix size is increased from 10 to 100 with same value of N, Data shows values for N=10.
Fig.3b. Comparison with average number of arithmetic operations per cell in existing algorithm

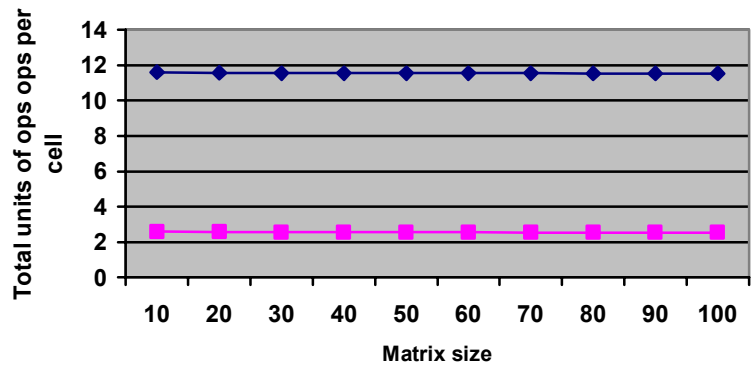

Fig.4. Comparison with the total operations plus overhead per cell in existing algorithm with same value of N, Data shows values for N=10

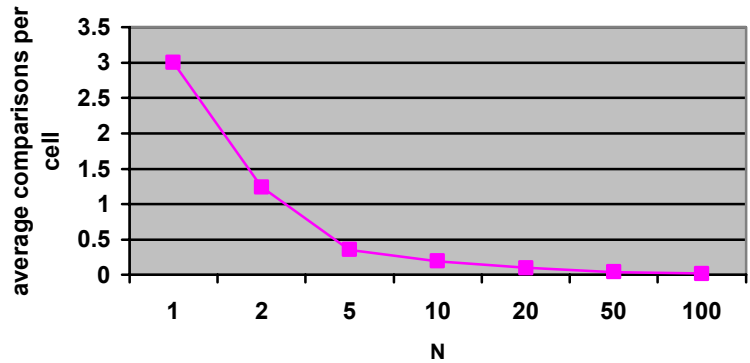

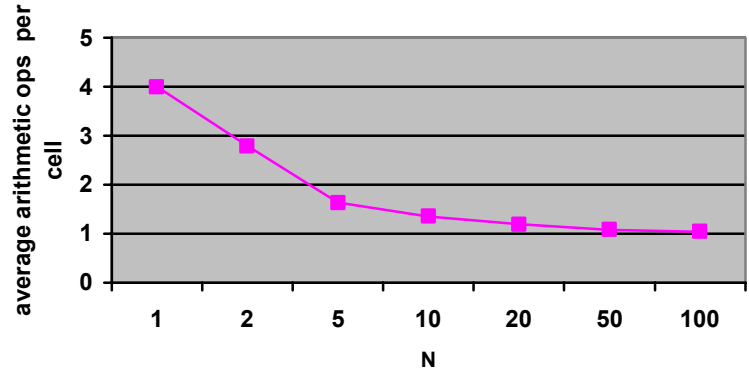

Fig.5a. Number of average comparisons required per cell if Value of N is increased from 1 to 100 with matrix size 100.
Fig.5b Number of average arithmetic operations required per cell if Value of N is increased from 1 to 100 with matrix size 100.

Now going ahead, check parameter is defined that makes the sequence to go through the gapped alignment after the middle path approach. As the name suggests, it is advised to be the middle path of cutoff for ungapped extension score and BLAST cutoff parameter for displaying a sequence.

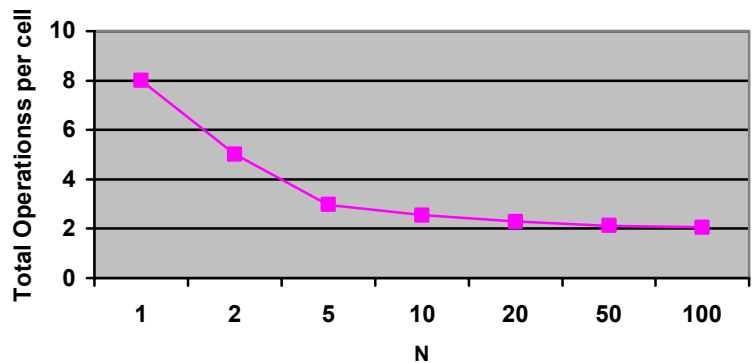

Fig.6. Number of Total operations required per cell if the Value of N is increased from 1 to 100 with matrix size 100

This parameter is called as $cutoff_{mp}$.
So As per BLAST parameters S1 and S2
$$cutoff_{mp} = (S1+S2)/2.$$
So if the score from the middle path approach is between $cutoff_{mp}$ and S2, then gapped alignment is performed as in BLAST; and if it is between S1 and $cutoff_{mp}$ then there is no need to perform the gapped alignment and directly go to the step for displaying the sequence. If it crosses S2, then also it can be skipped and not to go for the gapped alignment and can display the sequence. So the time saved is proportional to the number of sequences that will not go for gapped alignment. As per the calculations of middle path approach are concerned; they are not more then the 30 percent of the calculations made for gapped alignment of same sequences. This is shown in terms of reduction in the number of comparisons. In the similar way, there is reduction in the arithmetic calculations depending on different cases of the algorithm.

## 4. Discussion

The results are based on the work carried on HP x1433AP Model with 2 GB RAM and 2.8 GHz processor. The environment was Linux red hat 8.0. The existing NCBI Blast parameters, constants, default values and flags for comparisons except for the addition of middle path algorithm were used. These were compared with our program having the middle path approach. Previously for Similar experiments Brenner [14] and Park [15] have preferred to use Structural classification of proteins. Sequences were chosen randomly and searched against the entire database. The product was tested for various values of N.

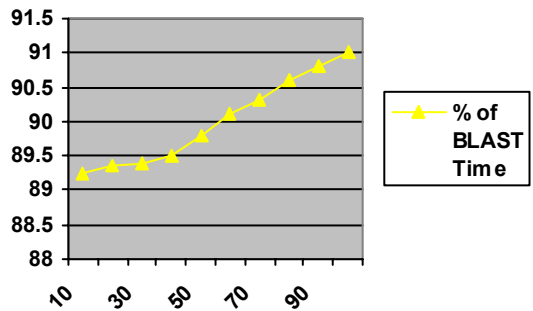

Fig.7. Comparison of the performance of new approach with the actual BLAST time with increasing value of N

As N is increased, the number of gapped alignments decreases that has direct impact on the execution time of BLAST. The total time by the gapped alignment is one third of the total time and lower bound for the number of alignments that will skip the gapped alignment test is 50% and time taken for additional calculations for remaining 50%.

Table2a. Tables show results of improvement in the algorithm and give percentage improvement when matrix size is increased and when value of N is unchanged.

| matrix size | Average Comparisons per cell for middle path approach | Average Comparisons per cell in the original algorithm | Average arithmetic operations per cell for middle path approach | Average arithmetic operations Per cell in the original algorithm | Total operations per cell for middle path approach | Total operations per cell in the original algorithm | % improvement |
|---|---|---|---|---|---|---|---|
| 10 | 0.21 | 5 | 1.39 | 5 | 2.6 | 11 | 76.37 |
| 20 | 0.2 | 5 | 1.37 | 5 | 2.57 | 11 | 76.64 |
| 30 | 0.197 | 5 | 1.363 | 5 | 2.56 | 11 | 76.73 |
| 40 | 0.195 | 5 | 1.36 | 5 | 2.555 | 11 | 76.78 |
| 50 | 0.194 | 5 | 1.358 | 5 | 2.552 | 11 | 76.8 |
| 60 | 0.193 | 5 | 1.355 | 5 | 2.548 | 11 | 76.84 |
| 70 | 0.1928 | 5 | 1.353 | 5 | 2.5458 | 11 | 76.86 |
| 80 | 0.1925 | 5 | 1.35 | 5 | 2.5425 | 11 | 76.89 |
| 90 | 0.1922 | 5 | 1.349 | 5 | 2.5412 | 11 | 76.90 |
| 100 | 0.192 | 5 | 1.349 | 5 | 2.54 | 11 | 76.91 |
| 1000 | 0.1901 | 5 | 1.3479 | 5 | 2.5380 | 11 | 76.93 |
| 10000 | 0.1897 | 5 | 1.3470 | 5 | 2.5367 | 11 | 76.94 |
| 100000 | 0.18903 | 5 | 1.34612 | 5 | 2.53515 | 11 | 76.96 |

Table2b. Tables show results of improvement in the algorithm and give percentage improvement when N is increased and matrix size remains unchanged.

| Value of N | Average Comparisons per cell | Average Comparisons per cell | Average arithmetic operations per cell | Average arithmetic operations Per cell | Total operations per cell | Total operations per cell | % improvement |
|---|---|---|---|---|---|---|---|
| 1 | 3 | 5 | 4 | 5 | 8 | 11 | 27.28 |
| 2 | 1.24 | 5 | 2.79 | 5 | 5.03 | 11 | 54.28 |
| 5 | 0.356 | 5 | 1.632 | 5 | 2.988 | 11 | 72.84 |
| 10 | 0.192 | 5 | 1.354 | 5 | 2.546 | 11 | 76.86 |
| 20 | 0.0985 | 5 | 1.189 | 5 | 2.2875 | 11 | 79.21 |
| 50 | 0.04 | 5 | 1.08 | 5 | 2.12 | 11 | 80.73 |
| 100 | 0.02 | 5 | 1.04 | 5 | 2.06 | 11 | 81.28 |

The experiments show that the ideal value for N is in between 9 to 11. The time improvement in such cases is approximately 75%. Out of these based on our cutoff$_{mp}$ at least half of the sequences will not go for gapped alignment (This is the lower bound and in some cases up to 81% of the sequences will not go for gapped alignment). By skipping half of sequences, savings are 37.5% of the time for gapped alignment. This includes the time taken to calculate the middle path approach to exclude the sequences for gapped alignment. The gapped alignment takes 33% of the total BLAST time so the total saving will be approximately 12% (37.5% of 33%). It is also noted that as the size of the sequences is increasing the method given here will be more useful as shown in the table with the calculations for sequences with matrix size 1000, 10000 and 100000. So as the matrix size id increasing there is no degradation of time component and there is no compromise on the sensitivity of the resulting sequences.

## 5. Conclusion

By introducing this middle path algorithm as part of BLAST and then instead of performing gapped alignment on all the sequences, a check called cutoff$_{mp}$ is performed, then only a very limited number of sequences will go for gapped alignment. This will result in performing less number of gapped alignments resulting in better speed and improving the overall performance of the BLAST algorithm. The scientific community using BLAST will be greatly benefited in their research and development in terms of saving the time in processing the genomic sequences.

## 6. Acknowledgements

This work was supported by All India Council of Technical Education (AICTE), Ministry of Human Resources and Development, Govt. of India, New Delhi by funding the project for "Efficient algorithm design for pattern discovery in bioinformatics sequences" which is currently being executed at Thapar Institute of Engineering and technology, Patiala, India.